\documentclass[12pt]{spieman}  
\usepackage{amsmath,amsfonts,amssymb}
\usepackage{graphicx}
\usepackage{setspace}
\usepackage{tocloft}

\title{Towards reliable calcification detection: calibration of uncertainty in object detection from coronary optical coherence tomography images}

\author[a]{Hongshan Liu}
\author[a]{Xueshen Li}
\author[b]{Abdul Latif Bamba}
\author[c]{Xiaoyu Song}
\author[d]{Brigitta C. Brott}
\author[d]{Silvio H. Litovsky}
\author[a,*]{Yu Gan}
\affil[a]{Stevens Institute of Technology, Biomedical Engineering, Hoboken, NJ, 07030}
\affil[b]{Columbia University, Department of Electrical Engineering, New York, NY, 10027}
\affil[c]{The Icahn School of Medicine at Mount Sinai, New York, NY, 10029}
\affil[d]{The University of Alabama at Birmingham, School of Medicine, Birmingham, AL, 35487}

\cftpagenumbersoff{figure}
\cftpagenumbersoff{table} 
\begin{document} 
\maketitle

\begin{abstract}

\textbf{Significance:} Optical coherence tomography (OCT) has become increasingly essential in assisting the treatment of coronary artery disease (CAD). However, unidentified calcified regions within a narrowed artery could impair the outcome of the treatment. Fast and objective identification is paramount to automatically procure accurate readings on calcifications within the artery. 

\textbf{Aim:} We aim to rapidly identify calcification in coronary OCT images using a bounding box and reduce the bias of predicting in automated prediction models.

\textbf{Approach:} We first adopt a deep learning-based object detection model to rapidly draw the calcified region from coronary OCT images using bounding box. We measure the uncertainty of predictions based on the expected calibration errors, thus assessing the certainty level of detection results. To calibrate confidence scores of predictions, we implement dependent logistic calibration using each detection result's confidence and center coordinates.

\textbf{Results:} We have developed an object detection module to draw the boundary of calcifided region at a rate of 140 frame per second. With the calibrated confidence score of each prediction, we lower the uncertainty of predictions in calcification detection and eliminate the estimation bias from various object detection methods. The calibrated confidence of prediction results in a confidence error of approximately 0.13, suggesting that the confidence calibration on calcification detection could provide a more trustworthy result.

\textbf{Conclusions:} Given the rapid detection and effective calibration of the proposed work, we expect it can assist clinical evaluation of treating the CAD during the imaging-guided procedure.

\end{abstract}


{\noindent \footnotesize\textbf{*}Yu Gan,  \linkable{ygan5@stevens.edu} }

\begin{spacing}{2}  

\section{Introduction}

Optical coherence tomography (OCT) can acquire high-resolution cross-sectional images of coronary arteries. The high quality and detailed information from coronary OCT images facilitates the treatment of coronary artery diseases (CAD). 
Coronary artery disease causes 1 of every 5 deaths in Europe \cite{nichols2013cardiovascular} and 1 of every 6 deaths in the US \cite{go2014heart}, and remains one of the leading causes of morbidity and mortality in developed countries \cite{okrainec2004coronary}. 
Coronary atherosclerosis is caused by the gradual buildup of plaque resulting from the depositing of calcium, lipids, and macrophages from the luminal blood into the arterial intima.
Coronary atherosclerosis compounds and augments the risks of heart attack and heart failure. 
When treated improperly or left unattended, coronary atherosclerosis blocks the pathways to the heart's main arteries, known as the coronary arteries. 
Potential effects of plaque in CAD include chest pain, shortness of breath, heart failure, myocardial infarction, and sudden death.

A typical treatment for CAD is percutaneous coronary intervention (PCI), which is a non-surgical procedure used to treat the narrowing of the heart's coronary arteries. 
Unidentified calcified tissues within a narrowing artery often negatively impact the benefits of treatment. Approximately 700,000 PCI are performed every year in the US, and calcifications have been found in 17\% to 35\% of patients undergoing the procedure \cite{moussa2005impact, farag2016treatment, kawaguchi2008impact}, highlighting a need to precisely locate the existence and extent of calcifications.
Most PCI procedures involve using stents to open up obstructed coronary arteries \cite{kim2014rate}.  During the PCI procedure, a catheter with a tiny, folded balloon on its tip is inserted into the blood vessels until it arrives at the site where the plaque buildup is causing a blockage. At that point, the balloon is inflated to compress the plaque against the artery walls, therefore widening the passageway, and restoring blood flow to the heart. After that, the balloon is deflated and removed. A stent implantation is performed in the plaque buildup area to keep the artery open after removing the balloon \cite{migliavacca2002mechanical}.
Excess coronary calcification is highly related to the suboptimal deployment of the stent in the coronary during the PCI \cite{sinitsyn2003relationships}. 
Major calcifications are of great concern for two reasons \cite{gharaibeh2019coronary}. Calcifications can lead to stent underexpansion and strut malapposition. Malapposition of stent struts (e.g. an empty space between the strut and the adjacent vessel wall) might preclude healthy endothelial tissue growth. 
Even though stent deployment is generally effective in the short-term, stent efficacy can be reduced and the risk can be increased by adverse clinical events, such as in-stent restenosis and thrombosis in the medium- and long-term \cite{edelman1998pathobiologic, dangas2010stent, fujii2005stent, attizzani2014mechanisms, doi2009impact, hong2006intravascular, uren2002predictors}.

Coronary imaging guidance during PCI is one of the key determinants of treatment outcomes. Imaging is integral to every stage of PCI, such as assessment of lesion severity, preprocedural planning, optimization, and management of immediate complications \cite{mehrotra2018imaging,richards2021vision}. 
Optical coherence tomography has significant advantages for characterizing coronary calcification that typically has a signal-poor area with sharply delineated borders \cite{fujino2018new}. A typical OCT system can achieve a high axial resolution at the micron level and a penetration depth of up to 2 mm, indicating superior imaging capability \cite{prati2012expert, bezerra2009intracoronary}.
Detection of calcified regions within coronary OCT images is critical for intervention \cite{wu2020automatic}. On account of this, developing an object detection algorithm that is capable of detecting calcification in OCT images is essential. 

Deep learning has been increasingly explored in analyzing the diseased tissue in coronary OCT images \cite{tong2020application}. Existing research work \cite{athanasiou2019deep,abdolmanafi2018characterization,lee2020segmentation,avital2021identification,shibutani2021automated,lee2019automated } have conducted extensive studies to automatically identify plaque in coronary OCT images. A weighted majority voting from different Convolutional Neural Networks (CNN) \cite{abdolmanafi2018characterization} was used to solve the multi-class classification problem of pathological formations in coronary artery tissues. A deep convolutional architecture named SegNet segmented calcification in coronary OCT images \cite{gharaibeh2019coronary}. A two-step deep learning approach \cite{lee2020segmentation} characterizes plaques in coronary arteries in OCT images, by first localizing the major calcification lesions using the CNN model and second applying the deep learning model (SegNet) to provide pixel-wise classifications of calcified plaques. A modified deep convolutional segmentation model UNet \cite{avital2021identification} was used to identify calcification in coronary OCT images. The segmentation module in MASK-RCNN is employed to identify erosion region \cite{Sun:22}. 
Currently, the most popular way to perform automated analysis on OCT images is deep learning-based segmentation, which makes the pixel-wise classification and outputs the detailed shape and location of tissue of interest. 
Demonstrably, segmentation architecture results in large computational costs due to burden from the pixel-wise classification. By virtue of this, a more efficient way of enacting automated analysis of coronary images is through the use of object detection, which outputs the bounding box of tissue region rather than pixel-wise classification of tissue region, in order to efficiently identify the diseased region in coronary images.

While existing works also focus more on increasing the accuracy of deep learning models, the quality of predictions can be negatively impacted by overconfident deep learning models \cite{ghoshal2022calibrated}. When an overconfident prediction is produced, the deep learning models tend to be excessively complex with respect to the structure of network. 
The overconfidence issue 
can lead to unreliable predictions that are biased by the specific network structure. 
In general, recalibration methods of the well-trained model, such as Platt scaling \cite{platt1999probabilistic}, histogram binning \cite{zadrozny2001obtaining}, temperature scaling \cite{guo2017calibration}, can improve the calibration of the overconfident prediction results. Besides, model ensemble methods \cite{tran2020hydra,wen2020batchensemble} can also reduce overconfidence by aggregating the prediction results over multiple models. 
However, there are limited studies on correcting overconfident predictions in coronary OCT images. In OCT-related CAD treatment, over-confidence could be dangerous, as confidence is often learned as the likelihood that the prediction is correct. 
Therefore, in safety-critical and risk-sensitive applications in clinical diagnosis, it is crucial to quantify and calibrate the uncertainty of predictions.

In this work, we aim to achieve reliable calcification detection for patients with CAD to boost the efficiency of clinical diagnosis. 
We summarize our contributions as:
\begin{enumerate}
    \item  We detect calcification in coronary OCT images via an object detection-based deep learning model. The object detection process delineates the bounding box of calcified region within OCT images, providing a computationally efficient solution in comparison with conventional segmentation methods. 
     \item We propose to calibrate the confidence of coronary object detection task. We use a dependent logistic calibration method to reduce the bias of uncertainty induced by the training of neural networks. 
     
      \item We quantitatively and qualitatively evaluate the effectiveness of the proposed work on a human coronary dataset. The experimental results demonstrate the accuracy and speed of calcification detection and also the effectiveness to reduce the bias of confidence among two most popular object detection methods.   
      
\end{enumerate}

\section{Methods}

\begin{figure}[h!]
\begin{center}
\includegraphics[width=\columnwidth]{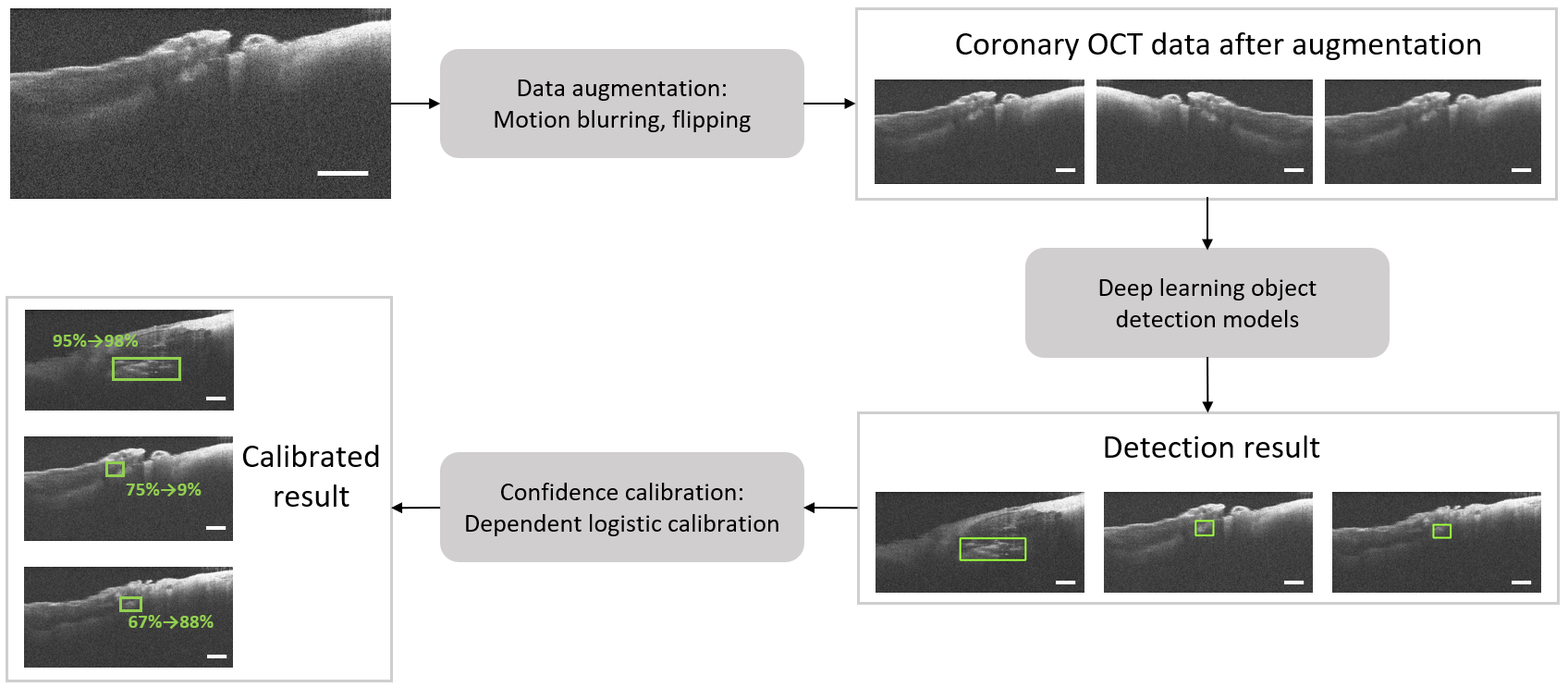}
\end{center}
\caption{Flowchart of the proposed work. Scale Bar: $500\mu m$ }\label{fig:4}
\end{figure}

The workflow is shown in Fig. \ref{fig:4}. The steps are: (1) The coronary OCT data is first processed by a data augmentation module to create motion-blurred and horizontally flipped copies of each original OCT image. (2) The coronary OCT data after augmentation is trained by deep learning object detection models and outputs the detection results on test data. (3) Detections containing bounding box coordinates and confidence scores are processed through dependent logistic calibration and output a calibrated confidence score for each predicted bounding box.

\subsection{Data collection}

Samples are imaged by the spectral domain OCT system (Thorlabs Ganymede, Newton, NJ) with an axial resolution of 3 $\mu$m and a lateral resolution of 4 $\mu$m in air. Autopsy specimens of human heart vessels are collected and imaged through the same protocol in \cite{23,li2022multi}. All images are acquired in the laboratory at the University of Alabama.  

\subsection{Data augmentation}
Various data augmentation techniques have been proposed to improve the performance of deep learning models \cite{hussain2017differential}.
In the time of imaging, the quality of OCT images may be impacted due to degradation caused by motion blur \cite{kraus2012motion,lian2018deblurring},  which appears when a camera is moving during image acquisition.
A motion blur filter is used to simulate this effect of real-world conditions \cite{abayomi2021cassava}. 
Other common augmentation strategies, such as flipping, cropping, scaling, Gaussian noise, rotation, and shears, are routinely performed \cite{sharif2014cnn}. 
Noticeably, we do not prefer vertical flipping and rotation in OCT images, because the light propagates in a fixed direction, applying such methods will change the nature of OCT images.

Therefore, in this work, two copies of each OCT image are created by applying a motion-blurring filter and flipping horizontally for training the deep learning model.


\subsection{Object detection}
\begin{figure}[h!]
\begin{center}
\includegraphics[width=\columnwidth]{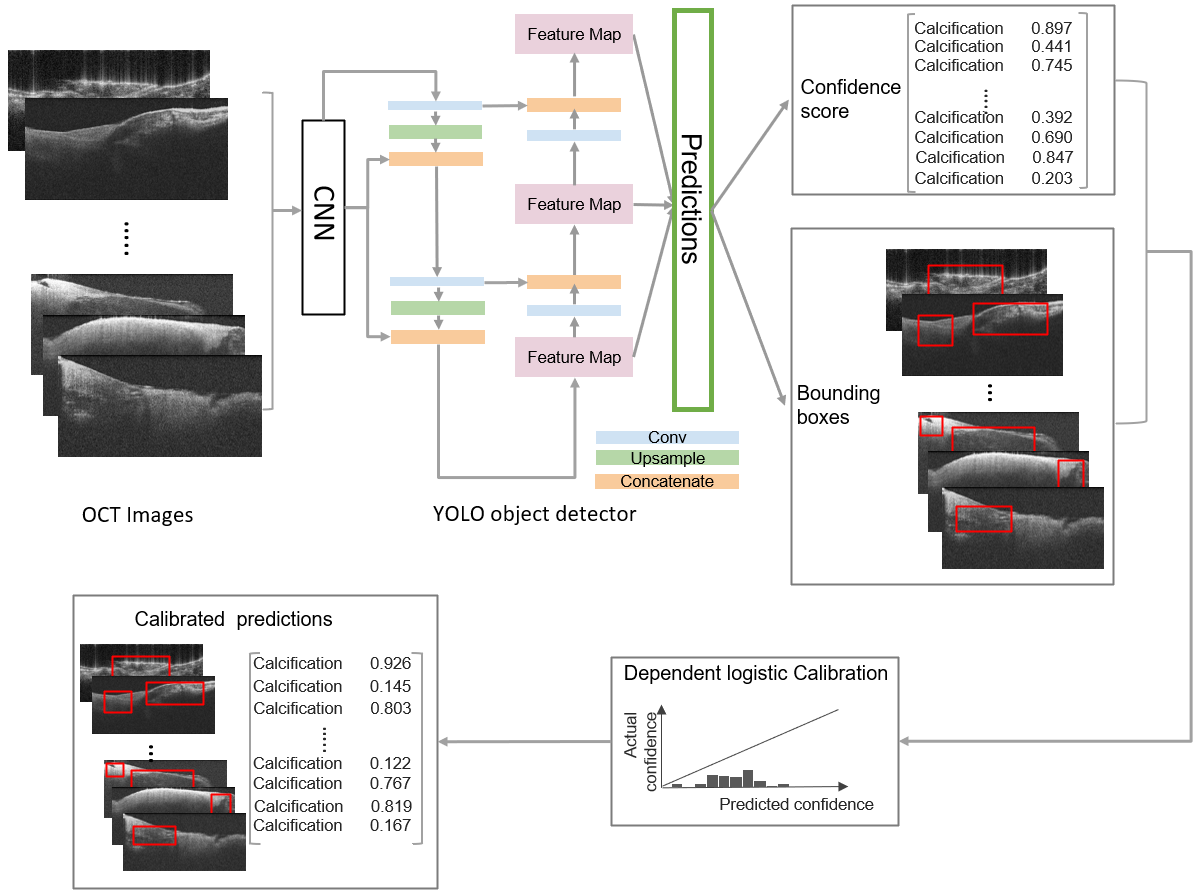}
\end{center}
\caption{The schematic of YOLO object detector and calibration.}\label{fig:yolo}
\end{figure}

Object detection creates bounding box regions that identify an object's position, size, and class within an image. We opt to use You-Only-Look-Once v5 (YOLO)\cite{glenn_jocher_2020_4154370} to rapidly identify the bounding box and tissue types within an OCT image. Because of its lightweight and feature-reuse properties, YOLO architecture is powerful to realize fast and accurate detection. 
As the schematic shown in Fig. \ref{fig:yolo}, YOLO enhances detection performance by utilizing different scales of feature maps to generate predictions for objects of different sizes.

The predictions have the outputs in two branches, confidence scores ($confidence$) and bounding boxes ($x_{center}$,$y_{center}$,$width$,$height$).
In the confidence score branch, the confidence score indicates a certain level that the prediction is true. 
In the bounding boxes branch, the values of the center coordinate, together with the width and height of the bounding box, depict the location of the predicted bounding box. 

\subsection{Uncertainty measurement and confidence calibration}
The common use of calibration is for the classification task, where only the confidence score can be utilized for a given image. In object detection, one additional piece of information that can be included for calibration is the location and scale of the bounding box. 
Therefore, in the object detection task, the criterion of a calibrated model is defined as the precision of a prediction given the confidence, class category, and bounding box information \cite{kuppers2020multivariate}, as in Eq. \ref{eq:01}: 

\begin{equation}
\begin{gathered}
P(correct\;prediction \;|\; p\hat = confidence, y\hat=y, r\hat=r)=confidence, \\
\forall conf \in [0,1], y\in \mathrm{Y}, r\in[0,1]^K 
\label{eq:01}
\end{gathered}
\end{equation}
where $y$ is the predicted class, and $r$ is the bounding box information with $k$ dimensions. 

Expected Calibration Error (ECE) is used to measure the uncertainty of the prediction of the deep learning model. The ECE of object detection is calculated by binning the confidence $\hat{p}$ into $M$ equally spaced bins. Samples with different confidence scores fall into corresponding bins. $B_m$ is the number of samples in a bin. The $Precision$ and $Confidence$ are the average precision and average confidence respectively among the samples in a bin. The ECE is given by:
\begin{equation}
ECE=\sum_{m=1}^{M}\frac{\left|B_m\right|}{N}|Precision\left(m\right)-Confidence\left(m\right)|               
\label{eq:02}
\end{equation}



For confidence calibration, in this work, we take two additional bounding box information, the center-x and center-y position along with the confidence score to calibrate the prediction results using the dependent logistic calibration \cite{kuppers2020multivariate}, where the multivariate probability density function is used to model the log-likelihood ratio (lr) of the combined input($confidence, bounding\;box$). Taking the correlations between confidence and the bounding box into consideration, the calibration map is defined as $g$:
\begin{equation}
g\left(input\right)\approx\frac{1}{1+e^{-lr(input)}},\;\;\;\;\;\;  		lr\left(s\right)=\frac{1}{2}[{(s}_-^T\mathrm{\Sigma}_-^{-1}s_-)-(s_+^T\mathrm{\Sigma}_+^{-1}s_+)]+c 
\label{eq:05}
\end{equation}
For the variable s, $s_{+} =s-\mu_{+}$ and $s_{-} =s-\mu_{-}$, and $c=\log|\frac{V_{-}}{V_{+}}|$, 
with ${\mu_+ }$ and ${\mu_-}$  as the mean vectors, ${\mathrm{\Sigma}_-}$ and ${\mathrm{\Sigma}_+}$ as the covariance matrices for incorrect and correct prediction respectively, for the true and false predictions. As shown in the calibrated predictions block in Fig. \ref{fig:yolo}, a new confidence score for each prediction is obtained by mapping the input to the calibration map $g$. The ECE of the prediction results before and after calibration is calculated to test the effect of calibration on model uncertainty.
 
\section{Results}

\subsection{Experimental setup}

For model development, we use 943 OCT images, from 14 OCT specimen segments, for a three-fold cross-validation. The OCT images were acquired from specimens that contain calcification regions, which include essential information for CAD treatments. Within each OCT volume, B-scans were sampled at an interval of 20. 
Each B-scan is with a size of 1024$\times$1500 pixels, corresponding to a space of 1.98$\times$3 mm$^2$. In the confidence calibration stage, 60\% predictions are used to fit the calibration model, with the rest 40\% predictions to be tested.


The YOLO was built in Python 3.8, PyTorch 1.10, CUDA 11.1, and NVIDIA RTX 6000, and a pre-trained weight \cite{yolo5s} was used in this work, with batch size of 8, the learning rate of 0.001 using Adam optimizer.
In addition to YOLO, another object detection scheme, Single Shot Multibox Detector (SSD) \cite{liu2016ssd}, was trained for comparison purposes.
The SSD was built in Python 3.8, PyTorch 1.10, CUDA 11.1, and NVIDIA RTX 6000, the training process started by loading pre-trained weight \cite{tensorflow_2021}, with batch size of 8, the learning rate of 0.001 using stochastic gradient descent optimizer with the momentum of 0.9.

\subsection{Object detection}

To evaluate the performance of calcification detection, three metrics, precision, recall, and f1-score, are calculated, as given in Eq.(\ref{eq:precision} $\sim$ \ref{eq:f1}). 
\begin{equation}
Precision = \frac{\# \; of \; true \;  positive}{\# \; of \; true \;positive + \# \;of\; false \;positive}
\label{eq:precision}
\end{equation}

\begin{equation}
Recall = \frac{\# \; of \; true \;  positive}{\# \; of \; true\; positive + \# \;of\; false \;negative}
\label{eq:recall}
\end{equation}

\begin{equation}
f1-score = \frac{2 \times precision \times recall}{precision + recall}
\label{eq:f1}
\end{equation}
where the true positive means the model correctly predicts the region with calcification, the false negative is the wrong prediction for the region that has calcification, and the false positive is the wrong prediction for the region with no calcification.
Precision indicates the number of correct predictions among all detection. Recall measures the fraction of correct predictions among ground truths. The f1-score is a measure of overall model performance by combining precision and recall.




\begin{figure}[h!]
\begin{center}
\includegraphics[width=\columnwidth]{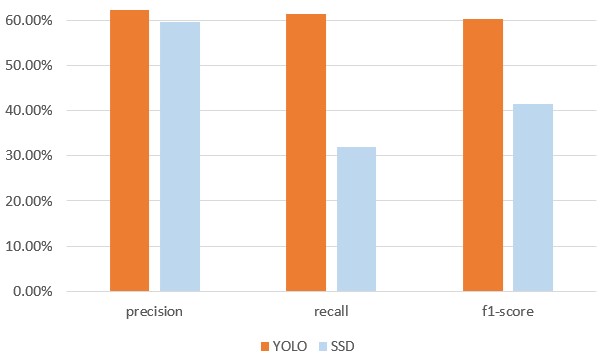}
\end{center}
\caption{ Object detection results of deep learning models in precision, recall, and f-1 score. The green bars are the results of YOLO, and the blue bars are the results of SSD.}\label{fig:detectionchart}
\end{figure}

\begin{figure}[ht]
\begin{center}
\includegraphics[width=\columnwidth]{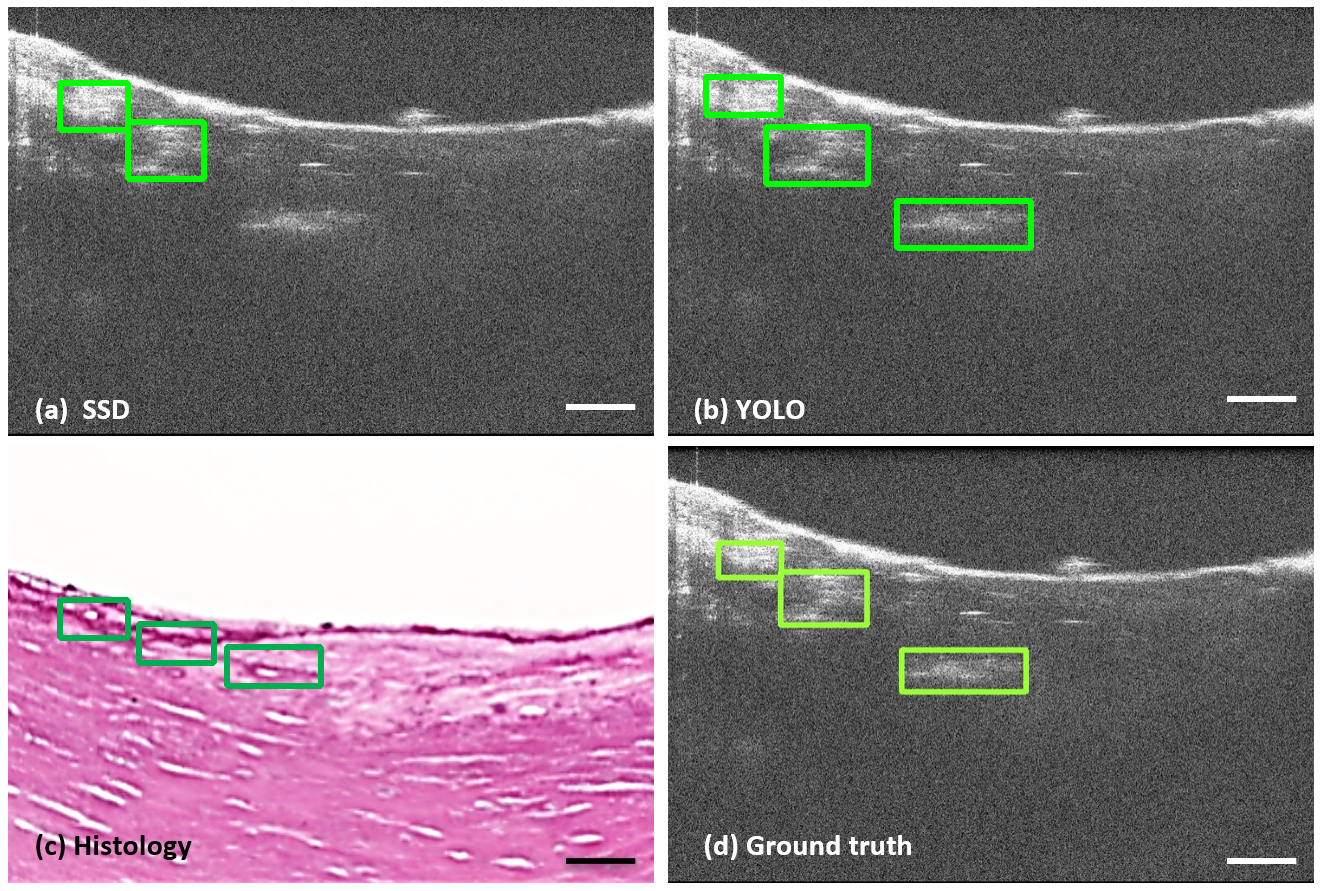}
\end{center}
\caption{Object detection results from (a) SSD, (b) YOLO. (c) corresponding histology, and (d) ground truth label. Scale Bar: $500\mu m$ }\label{fig:1}
\end{figure}

The quantitative results in Fig.\ref{fig:detectionchart} indicate that YOLO generally outperforms SSD in detecting calcification in coronary OCT images. 
YOLO outperforms SSD by  $\sim 5\%$ in precision, $\sim 30\%$ in recall, and $\sim 19\%$ in f1-score. The lower recall of SSD reveals higher false negative predictions, which agrees with the observation in Fig. \ref{fig:1}, where YOLO predicts all calcification in this coronary OCT image while SSD fails to detect the calcification region in relatively lower contrast.


Besides, as shown in table \ref{table:runtime}, the processing speed of YOLO is 140 frames per second (fps), while SSD processes at 68 fps, showing that YOLO has great capability for real-time detection, which is especially desirable in the circumstance of processing the large volume of OCT images. Comparatively, the runtime of OCT segmentation is approximately 15 fps \cite{he2018topology,shah2018multiple}, which indicates a larger computational burden than detection.

\begin{table}[]
\centering
\begin{tabular}{l|cc}
              & YOLO  & SSD \\ \hline
runtime (fps) & 140          & 68
\end{tabular}
\caption{Runtime in frames per second (fps) for deep learning models detecting calcification in coronary OCT images in this work. }
\label{table:runtime}
\end{table}


\subsection{Uncertainty measurement and confidence calibration} 

\begin{figure}[h!]
\begin{center}
\includegraphics[width=\columnwidth]{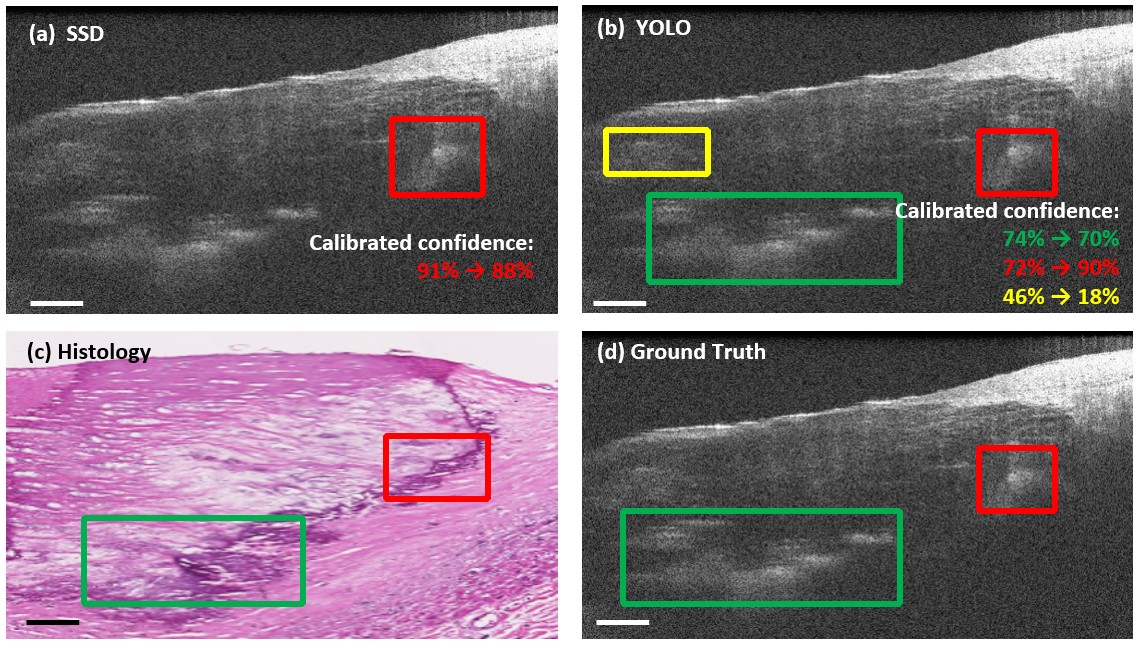}
\end{center}
\caption{Confidence calibration results of YOLO. (a) YOLO predictions, (b) ground truth labels, and (c) corresponding histology.  Scale Bar: $500\mu m$ }\label{fig:6}
\end{figure}

We evaluate the effectiveness of the calibration of predictions from both deep learning models. In Fig. \ref{fig:6}, the adjustment of confidence scores is observed in the calibrated predictions in YOLO and SSD. In Fig. \ref{fig:6}a, the prediction from SSD in the red box gets the confidence score slightly adjusted from 91\% to 88\%. In Fig. \ref{fig:6}b, the overconfident predictions shown in the yellow boxes reduced the confidence score from 46\% to 18\% after calibration, while the other confidence scores of predicted boxes are slightly adjusted.

\begin{table}[]
\centering
\begin{tabular}{c|ccc}
                         & YOLO  & SSD   & Difference \\ \hline
ECE                      & 0.233 & 0.731 & 0.498    \\
Calibrated ECE           & 0.134 & 0.146 & 0.008    \\
Before/After Calibration & 0.099 & 0.585 &    ——         


\end{tabular}
\caption{\centering The uncertainty measurements before and after confidence calibration: Expected Calibration Error (ECE), the last column shows the difference (SSD-YOLO) between two deep learning models. The rows of Before/After Calibration shows the changes in ECE during the calibration.}
\label{table:cali}
\end{table}

For quantitative evaluation, we use the ECE to measure the uncertainty of predictions. In Table \ref{table:cali}, before calibration, YOLO has a lower level of uncertainty in ECE, indicating that YOLO produces more reliable predictions. For both YOLO and SSD, the calibration errors are lowered to the same level around $\sim 0.14$ after implementing calibration, which proves that the calibration process can reduce the bias from different deep learning models and rectify the overconfident predictions of detectors.

\section{Discussion and conclusion}

In this work, we report calcification detection in coronary OCT images using deep learning models with uncertainty measurements and confidence calibration to reduce the bias in deep learning models. Although tissue detection and segmentation in OCT images have been studied, to our best knowledge, this work is the first to implement uncertainty measurement and confidence calibration for deep learning-based calcification detection in coronary OCT images. We investigate and compare the calcification detection performance of YOLO with SSD, and prove that YOLO is superior to SSD in producing more accurate and reliable predictions by detection accuracy and uncertainty measures. With an exceptional runtime of 140 fps, YOLO has the potential to become the real-time detector for predicting calcification in coronary OCT images. This work also implements confidence calibration by integrating the bounding box information with the confidence score. The quantitative and qualitative results show the effectiveness of the calibration, indicating its practical value in safe-critical and risk-sensitive applications, for example, the calcification detection in coronary OCT images during PCI.

In the future, we will implement other calibration methods on the predicted confidence score and seek to ensemble multiple models in order to produce more robust and reliable predictions for calcification detection in OCT images. Furthermore, by providing additional information critical to diagnosis, the calibrated confidence and uncertainty measures can be used in future clinical practice.

\section*{Funding}
This work was supported in part by National Science Foundation
(CRII-1948540), New Jersey Health Foundation, the National Center for Advancing Translational Research of the National Institutes of Health underaward number UL1TR003096.

\section*{Disclosures}
All authors declare that they have no conflicts of interest.

\section*{Acknowledgements}
The authors would like to thank Dr. Dezhi Wang from the University of Alabama, Birmingham, for histology service.

\section*{Data availability statement}
The datasets generated and analyzed in this work are available from the corresponding author upon reasonable request.





\bibliography{sample}   
\bibliographystyle{spiejour}   


\vspace{2ex}\noindent\textbf{Hongshan Liu} received her M.S. degree in Electrical Engineering from University of Michigan-Ann Arbor and her B.S. degree in Physics from Zhejiang University. She is a doctoral student in Biomedical Engineering at Stevens Institute of Technology. Her research focuses on deep learning-based image processing in the clinical applications of optical coherence tomography. 

\vspace{2ex}\noindent\textbf{Xueshen Li} received his M.S. degree in Biomedical Engineering from Eindhoven University of Technology. He is a doctoral student in Biomedical Engineering department at Stevens Institute of Technology. His research focuses on biomedical image processing.

\vspace{2ex}\noindent\textbf{Abdul Latif Bamba} is an undergraduate student from Columbia University. His research is in image processing, microelectronic and solid-state devices.

\vspace{2ex}\noindent\textbf{Xiaoyu Song} received her DrPH in Biostatistics from Columbia University. She is an Assistant Professor at the Icahn School of Medicine at Mount Sinai. Her research interest is in biostatistics and statistical genomics.

\vspace{2ex}\noindent\textbf{Brigitta C. Brott}, MD is an interventional cardiologist with a background in Materials Science and Engineering.  She obtained her cardiology and interventional cardiology training at Duke University Medical Center.  She is a Professor of Medicine and Biomedical Engineering at the University of Alabama at Birmingham.  Her research interests include novel coatings to improve healing after device implantation, and optimization of imaging and physiology assessments to guide cardiac interventional procedures.

\vspace{2ex}\noindent\textbf{Silvio H. Litovsky} received his MD degree from the University of Buenos Aires, Argentina. He is Professor of Pathology at the University of Alabama at Birmingham. His research interest includes multiple areas of cardiovascular pathology including high risk atherosclerotic plaques.  

\vspace{2ex}\noindent\textbf{Yu Gan} received his PhD degree in Electrical Engineering from Columbia University. He is an Assistant Professor in Biomedical Engineering at Stevens Institute of Technology. His research interest is in optical coherence tomography, biomedical image processing, and computer vision.



\end{spacing}
\end{document}